\documentclass[prd,twocolumn,twoside,preprintnumbers,superscriptaddress,nofootinbib]{revtex4}
\usepackage[a4paper, hdivide={1.91cm,,1.165cm}, vdivide={1.83cm,,3.2cm}]{geometry}

\usepackage{amsmath}
\usepackage{graphicx}
\usepackage{dcolumn}
\usepackage[hyperfootnotes=false]{hyperref}
\usepackage{xspace}
\usepackage{color}

\newcommand{\eq}[1]{Eq.~\eqref{#1}}

\newcommand{\real}{\mathrm{Re}\,}

\newcommand{\MeV}{\,\text{MeV}}
\newcommand{\GeV}{\,\text{GeV}}
\newcommand{\TeV}{\,\text{TeV}}

\newcommand{\BKs}{$B\to K^*\mu^+\mu^-$\ }

\begin{document}
\preprint{\vbox{\hbox{CERN-PH-TH-2015-001}\hbox{ULB-TH/14-26}}}

\title{Explaining $h\to\mu^\pm\tau^\mp$, $B\to K^* \mu^+\mu^-$ and $B\to K \mu^+\mu^-/B\to K e^+e^-$\\ in a two-Higgs-doublet model with gauged $L_\mu-L_\tau$}

\author{Andreas Crivellin}
\affiliation{CERN Theory Division, CH--1211 Geneva 23, Switzerland}%
\author{Giancarlo D'Ambrosio}
\affiliation{CERN Theory Division, CH--1211 Geneva 23, Switzerland}%
\affiliation{INFN-Sezione di Napoli, Via Cintia, 80126 Napoli, Italy}
\author{Julian Heeck}
\affiliation{Service de Physique Th\'eorique, Universit\'e Libre de Bruxelles, Boulevard du Triomphe, CP225, 1050 Brussels, Belgium}

\begin{abstract}

The LHC observed so far three deviations from the Standard Model (SM) predictions in flavour observables: LHCb reported anomalies in $B\to K^* \mu^+\mu^-$ and $R(K)=B\to K \mu^+\mu^-/B\to K e^+e^-$ while CMS found an excess in $h\to\mu\tau$. We show, for the first time, how these deviations from the SM can be explained within a single well-motivated model: a two-Higgs-doublet model with gauged $L_\mu-L_\tau$ symmetry. We find that, despite the constraints from $\tau\to\mu\mu\mu$ and $B_s$--$\overline{B}_s$ mixing, one can explain $h \to\mu\tau$, $B\to K^* \mu^+\mu^-$ and $R(K)$ simultaneously, obtaining interesting correlations among the observables.

\end{abstract}
%

\maketitle

\section{Introduction}
\label{intro}

So far, the LHC completed the SM by discovering the last missing piece, the Brout--Englert--Higgs particle~\cite{Aad:2012tfa,Chatrchyan:2012ufa}. Furthermore, no significant direct evidence for physics beyond the SM has been found, i.e.~no new particles were discovered. However, the LHC did observe three 'hints' for new physics (NP) in the flavor sector, which are sensitive to virtual effects of new particles and can be used as guidelines towards specific NP models: $h\to\mu\tau$, $B\to K^* \mu^+\mu^-$, and $R(K)=B\to K \mu^+\mu^-/B\to K e^+e^-$. It is therefore interesting to examine if a specific NP model can explain these three anomalies simultaneously, predicting correlations among them. 

LHCb reported deviations from the SM predictions~\cite{Egede:2008uy,Altmannshofer:2008dz} (mainly in an angular observable called $P_5^\prime$~\cite{Descotes-Genon:2013vna}) in $B\to K^* \mu^+\mu^-$~\cite{Aaij:2013qta} with a significance of $2$--$3\,\sigma$ depending on the assumptions of hadronic uncertainties~\cite{Descotes-Genon:2014uoa,Altmannshofer:2014rta,Jager:2014rwa}. This discrepancy can be explained in a model independent approach by rather large contributions to the Wilson coefficient $C_9$~\cite{Descotes-Genon:2013wba,Altmannshofer:2013foa,Horgan:2013pva}, i.e.~an operator $(\overline{s}\gamma_\alpha P_L b)(\overline{\mu}\gamma^\alpha \mu)$, which can be achieved in models with an additional heavy neutral $Z^\prime$ gauge boson~\cite{Gauld:2013qba,Buras:2013dea,Altmannshofer:2014cfa}. Furthermore, LHCb~\cite{Aaij:2014ora} recently found indications for the violation of lepton flavour universality in
\begin{equation}
	R(K)=\frac{B\to K \mu^+\mu^-}{B\to K e^+e^-}=0.745^{+0.090}_{-0.074}\pm 0.036\,,
\end{equation}
which disagrees from the theoretically rather clean SM prediction $R_K^{\rm SM}=1.0003 \pm 0.0001$~\cite{Bobeth:2007dw} by $2.6\,\sigma$. A possible explanation is again a NP contributing to $C_9^{\mu\mu}$ involving muons, but not electrons~\cite{Alonso:2014csa, Hiller:2014yaa,Ghosh:2014awa}. Interestingly, the value for $C_9$ required to explain $R(K)$ is of the same order as the one required by $B\to K^* \mu^+\mu^-$ \cite{Hurth:2014vma,Altmannshofer:2014rta}. In Ref.~\cite{Altmannshofer:2014cfa}, a model with gauged muon minus tauon number ($L_\mu-L_\tau$) was proposed in order to explain the \BKs anomaly. 

Concerning Higgs decays, CMS recently measured a lepton-flavour violating (LFV) channel~\cite{CMS:2014hha} 
\begin{equation}
	{\rm Br} [h\to\mu\tau] = \left( 0.89_{-0.37}^{+0.40} \right)\% \,,
	\label{h0taumuExp}
\end{equation}
which disagrees from the SM (where this decay is forbidden) by about $2.4 \,\sigma$. 
Such LFV SM Higgs couplings are induced by a single operator up to dim-6 and ${\rm Br}[h\to\mu\tau]$ can easily be up to $10\%$ taking into account this operator only~\cite{Harnik:2012pb,Blankenburg:2012ex,Davidson:2012ds,Arhrib:2012ax,Arhrib:2012mg,Kopp:2014rva}. However, it is in general difficult to get dominant contributions to this operator in a UV complete model, as for example in models with vector-like leptons \cite{Falkowski:2013jya}. 
Therefore, among the several attempts to explain this $h\to\mu\tau$ observation~\cite{Dery:2014kxa,Campos:2014zaa,Celis:2014roa,Sierra:2014nqa,Lee:2014rba}, most of them are relying on models with extended Higgs sectors. One solution employs a two-Higgs-doublet model (2HDM) with gauged $L_\mu-L_\tau$~\cite{Heeck:2014qea}.

The abelian symmetry $U(1)_{L_\mu-L_\tau}$ is interesting in general: not only is this an anomaly-free global symmetry within the SM~\cite{He:1990pn,Foot:1990mn,He:1991qd}, it is also a good zeroth-order approximation for neutrino mixing with a quasi-degenerate mass spectrum, predicting a maximal atmospheric and vanishing reactor neutrino mixing angle~\cite{Binetruy:1996cs,Bell:2000vh,Choubey:2004hn}. Breaking $L_\mu-L_\tau$ is mandatory for a realistic neutrino sector, and such a breaking can also induce charged LFV processes, such as $\tau\to3\mu$~\cite{Dutta:1994dx,Heeck:2011wj} and $h\to \mu \tau$~\cite{Heeck:2014qea}.

Supplementing the model of Ref.~\cite{Heeck:2014qea} with the induced $Z'$ quark couplings of Ref.~\cite{Altmannshofer:2014cfa} can resolve all three anomalies from above. Interestingly, the semileptonic $B$ decays imply lower limit on $g^\prime/M_{Z^\prime}$, which allows us to set a lower limit on $\tau\to\mu\mu\mu$, depending on $h\to\mu\tau$.

\section{The model}

Our model under consideration is a 2HDM with a gauged $U(1)_{L_\mu-L_\tau}$ symmetry~\cite{Heeck:2014qea}. The $L_\mu-L_\tau$ symmetry with the gauge coupling $g^\prime$ is broken spontaneously by the vacuum expectation value (VEV) of a scalar $\Phi$ with $Q^{\Phi}_{L_\mu-L_\tau}=1$, leading to the $Z'$ mass
\begin{align}
m_{Z'} = \sqrt2 g' \langle\Phi\rangle \equiv g' v_\Phi\,,
\end{align}
and Majorana masses for the right-handed neutrinos\footnote{Active neutrino masses are generated via seesaw with close-to-maximal atmospheric mixing and quasi-degenerate masses~\cite{Heeck:2014qea}.}. 

Two Higgs doublets are introduced which break the electroweak symmetry: $\Psi_1$ with $Q^{\Psi_1}_{L_\mu-L_\tau}=-2$ and $\Psi_2$ with $Q^{\Psi_2}_{L_\mu-L_\tau}=0$. Therefore, $\Psi_2$ gives masses to quarks and leptons while $\Psi_1$ couples only off-diagonally to $\tau\mu$:
\begin{align}
\begin{split}
\mathcal{L}_Y \ \supset\ &-\overline{\ell}_f Y^\ell_{i}\delta_{fi} \Psi_2 e_i - \xi_{\tau\mu} \overline{\ell}_3 \Psi_1 e_2 \\
& -\overline{Q}_f Y^u_{fi} \tilde{\Psi}_2 u_i - \overline{Q}_f Y^d_{fi} \Psi_2 d_i + \mathrm{h.c.}\,.
\end{split}
\label{eq:yukawas}
\end{align}
Here $Q$ ($\ell$) is the left-handed quark (lepton) doublet, $u$ ($e$) is the right-handed up-quark (charged-lepton) and $d$ the right-handed down quark while $i$ and $f$ label the three generations. The scalar potential is the one of a $U(1)$-invariant 2HDM~\cite{Branco:2011iw} with additional couplings to the SM-singlet $\Phi$, which most importantly generates the doublet-mixing term 
\begin{equation}
V(\Psi_1,\Psi_2,\Phi)\ \supset \ 2 \lambda \Phi^2 \Psi_2^\dagger \Psi_1 \to \lambda v_\Phi^2 \Psi_2^\dagger \Psi_1 \equiv m_3^2 \Psi_2^\dagger \Psi_1\,,\nonumber
\end{equation}
that induces a small vacuum expectation value for $\Psi_1$~\cite{Heeck:2014qea}. We define $\tan\beta = \langle \Psi_2\rangle/\langle \Psi_1\rangle$ and $\alpha$ is the usual mixing angle between the neutral CP-even components of $\Psi_1$ and $\Psi_2$ (see for example~\cite{Branco:2011iw}). We neglect the additional mixing of the CP-even scalars with Re$[\Phi]$.

Quarks and gauge bosons have standard type-I 2HDM couplings to the scalars. The only deviations are in the lepton sector: while the Yukawa couplings $Y^\ell_{i}\delta_{fi}$ of $\Psi_2$ are forced to be diagonal due to the ${L_\mu-L_\tau}$ symmetry, $\xi_{\tau\mu}$ gives rise to an off-diagonal entry in the lepton mass matrix:
\begin{equation}
m^\ell_{fi}= \frac{v}{\sqrt{2}}\begin{pmatrix}
y_e\sin\beta &0&0\\
0& y_\mu \sin\beta& 0\\
0&\xi_{\tau\mu} \cos\beta& y_\tau \sin\beta
\end{pmatrix} .
\end{equation}
It is this $\tau$--$\mu$ entry that leads to the LFV couplings of $h$ and $Z'$ of interest to this letter. The lepton mass basis is obtained by simple rotations of $(\mu_R,\tau_R)$ and $(\mu_L,\tau_L)$ with the angles $\theta_R$ and $\theta_L$, respectively:
\begin{align}
 \sin \theta_R \simeq \frac{v}{\sqrt{2} m_\tau} \xi_{\tau\mu} \cos\beta \,, && \frac{\tan\theta_L}{\tan\theta_R} = \frac{m_\mu}{m_\tau} \ll 1\,.
\label{eq:thetaR}
\end{align} 
The angle $\theta_L$ is automatically small and will be neglected in the following.\footnote{Choosing $Q_{L_\mu-L_\tau}=+2$ for $\Psi_2$ would essentially exchange $\theta_L \leftrightarrow \theta_R$~\cite{Heeck:2014qea}, with little impact on our study.}
A non-vanishing angle $\theta_R$ not only gives rise to the LFV decay $h\to\mu\tau$ due to the coupling
\begin{equation}
\frac{m_\tau}{v}\frac{\cos(\alpha-\beta)}{\cos(\beta)\sin(\beta)}\sin(\theta_R)\cos(\theta_R) \bar\tau P_R\mu h\equiv \Gamma^{h}_{\tau\mu}\bar\tau P_R\mu h\,,\label{h0taumu}
\end{equation}
in the Lagrangian, but also leads to off-diagonal $Z'$ couplings to right-handed leptons
\begin{align}
g^\prime Z^\prime_\nu \, (\overline{\mu}, \overline{\tau})
\begin{pmatrix}
 \cos 2\theta_R& \sin 2\theta_R\\
\sin 2\theta_R& - \cos 2\theta_R
\end{pmatrix} \gamma^\nu P_R 
\begin{pmatrix}
 \mu\\
\tau
\end{pmatrix} ,
\end{align}
while the left-handed couplings are to a good approximation flavour conserving. In order to explain the observed anomalies in the $B$ meson decays, a coupling of the $Z'$ to quarks is required as well, not inherently part of $L_\mu-L_\tau$ models (aside from the kinetic $Z$--$Z'$ mixing, which is assumed to be small). Following Ref.~\cite{Altmannshofer:2014cfa}, we introduce heavy vector-like quarks, i.e.~$Q_L \equiv (U_L,D_L)$, $D_R^c$, $U_R^c$ and their chiral partners $\tilde{Q}_R \equiv (\tilde{U}_R, \tilde{D}_R)$, $\tilde{D}_L^c$, $\tilde{U}_L^c$, with vector-like mass terms
\begin{align}
 m_Q \overline{Q}_L \tilde{Q}_R + m_D \overline{\tilde{D}}_L D_R + m_U \overline{\tilde{U}}_L U_R + \mathrm{h.c.}\,,
\end{align}
and $L_\mu-L_\tau$ charges $+1$ (i.e.~$Q^{D_R}_{L_\mu-L_\tau}=Q^{U_R}_{L_\mu-L_\tau}=-1$), coupling them to the $Z^\prime$ boson. Yukawa-like couplings involving the heavy vector-quarks, the light chiral quarks and $\Phi$
\begin{align}
&\Phi \sum_{j = 1}^3 \left( \overline{\tilde{D}}_R  Y^Q_{j} P_L d_j +\overline{\tilde{U}}_R  Y^Q_{j} P_L u_j \right)\\
& + \Phi^\dagger \sum_{j = 1}^3 \left( \overline{\tilde{D}}_L  Y^D_{j} P_R d_j +\overline{\tilde{U}}_L Y^U_{j} P_R u_j \right)  + \mathrm{h.c.}\nonumber
\end{align}
then induce couplings of the SM quarks to the $Z'$ once $\Phi$ acquires its VEV. Thus, integrating out the heavy vector-like quarks gives rise to effective $Z^\prime\bar d_i d_j$ couplings \cite{Langacker:2008yv,Buras:2012jb} of the form
\begin{equation}
g'\left({{{\bar d}}_i}{\gamma ^\mu }{P_L}{{d}_j}{Z'_\mu }\Gamma_{ij}^{{d}L} + {{{\bar d}}}_i{\gamma ^\mu }{P_R}{{d}_j}{Z'_\mu }\Gamma_{ij}^{{d}R}\right) ,
\end{equation}
with hermitian matrices $\Gamma_{ij}^{{d}L}$ that are related to the vector quark masses $m_{Q,D,U}$ and Yukawa couplings $Y^{Q,D,U}$ as follows~\cite{Altmannshofer:2014cfa}:
\begin{align}
\Gamma_{ij}^{d R} \simeq -\frac{v_\Phi^2}{2 m_D^2} (Y^D_{i} Y^{D*}_{j}) \,, &&
\Gamma_{ij}^{d L} \simeq \frac{v_\Phi^2}{2 m_Q^2} (Y^Q_{i} Y^{Q*}_{j}) \,,
\label{eq:quarkcouplings}
\end{align}
which holds in the approximation $|\Gamma_{ij}^{q R/L}| \ll 1 $.\footnote{Compared to Ref.~\cite{Altmannshofer:2014cfa}, the vector-like quarks also have Yukawa couplings $y_{\Psi_1}$ to the $L_\mu-L_\tau$-charged scalar doublet $\Psi_1$. This induces a small additional mass mixing among the heavy quarks, and also a coupling to $h$ suppressed by $y_{\Psi_1} \cos (\alpha-\beta)$. We assume these couplings $y_{\Psi_1}$ to be small to avoid large contributions to $gg\to h$ and $h\to \gamma\gamma$.}

\section{Flavour observables}
\label{flavour}

We will now recall the necessary formula in the region of interest (i.e.~small $\theta_R$) considering only the processes giving to most relevant bounds on our model, i.e.~$B_s$--$\bar B_s$ mixing, neutrino trident production and $\tau\to3\mu$. 

\begin{figure*}
\includegraphics[width=0.43\textwidth]{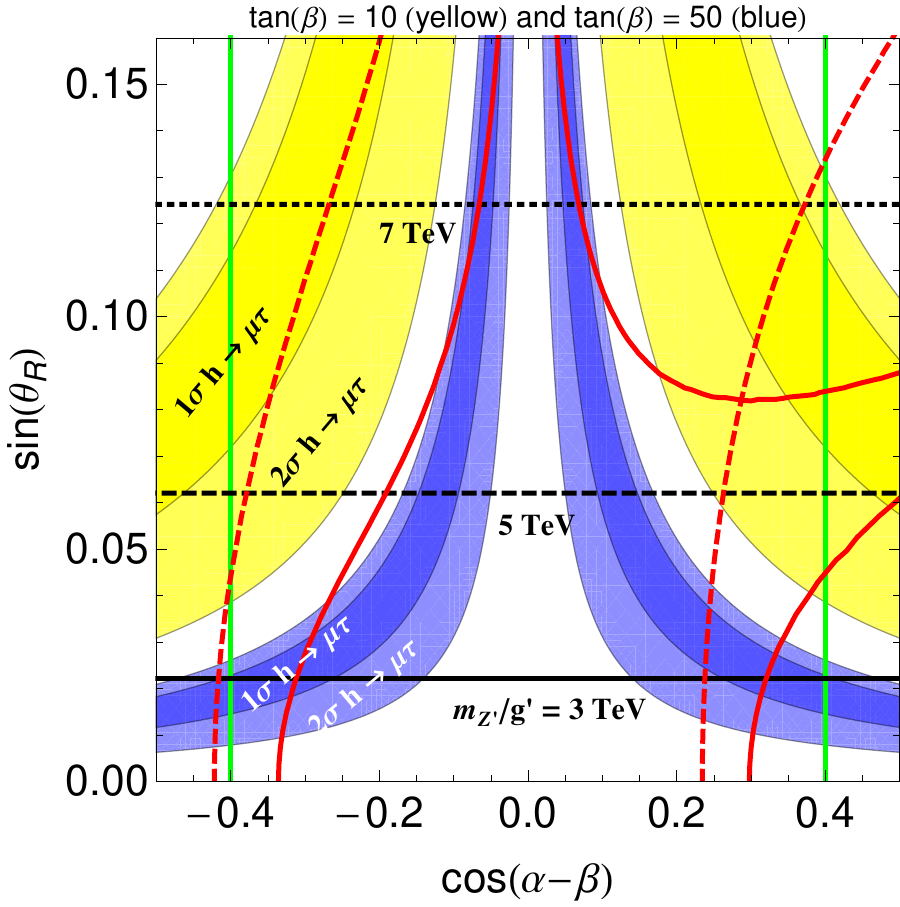} \hspace{3ex}
\includegraphics[width=0.42\textwidth]{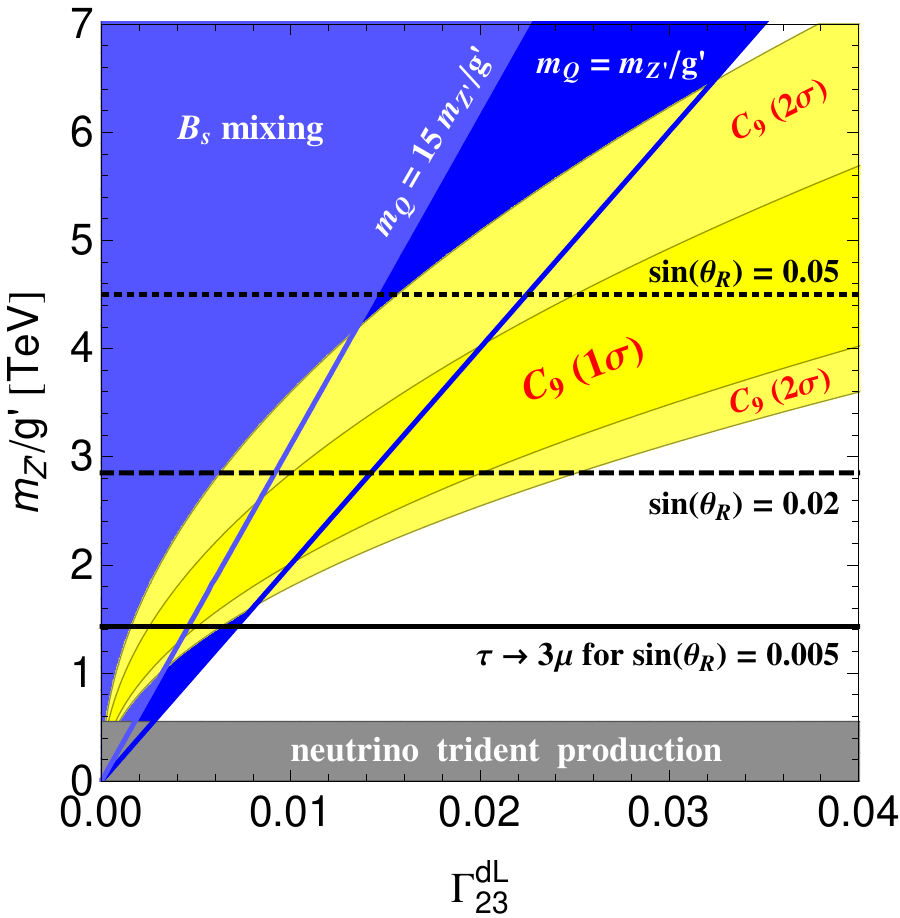}
\caption{Left: Allowed regions in the $\cos(\alpha-\beta)$--$\sin(\theta_R)$ plane. The blue (light blue) region corresponds to the $1\sigma$ ($2\sigma$) region of the CMS measurement of $h\to\mu\tau$ for $\tan \beta = 50$; yellow stands for $\tan \beta=10$. The (dashed) red contours mark deviations of $h\to\tau\tau$ by $10\%$ compared to the SM for $\tan\beta = 50$ ($10$). The vertical green lines illustrate the naive LHC limit $|\cos(\alpha-\beta)| \lesssim 0.4$, horizontal lines denote the $90\%$~C.L.~limit on $\tau\to 3\mu$ via $Z'$ exchange. \\
Right: Allowed regions in the $\Gamma^{dL}_{23}$--$m_{Z'}/g'$ plane from \BKs and $R(K)$ (yellow) and $B_s$ mixing (blue). For $B_s$ mixing (light) blue corresponds to ($m_Q=15 m_{Z'}/g'$) $m_Q=m_{Z'}/g'$. The horizontal lines denote the lower bounds on $m_{Z'}/g'$ from $\tau\to3\mu$ for $\sin(\theta_R)=0.05,\; 0.02,\; 0.005$. The gray region is excluded by NTP.}
\label{fig:HiggsPlot}
\end{figure*}

\subsection{$h\to\mu\tau$}

The branching ratio for $h\to\mu\tau$ reads
\begin{equation}
{\rm Br}\left[h\to\mu\tau\right]\simeq \dfrac{m_{h}}{8\pi \Gamma_{\rm SM}} \left| {\Gamma^{h}_{\tau\mu}} \right|^2 ,
\end{equation}
where $\Gamma_{\rm SM}\simeq 4.1 \MeV$ is the decay width in the SM for a 125~GeV Higgs~\cite{Dittmaier:2012vm} and $\Gamma^{h}_{\tau\mu}$ is defined in \eq{h0taumu}. Comparing this to \eq{h0taumuExp} one sees that both $\sin\theta_R\neq0$ and $\cos(\alpha-\beta)\neq0$ are required to explain the CMS excess~\cite{Heeck:2014qea}.

\subsection{Lepton decays}

While the Higgs contributions to $\tau\to\mu\mu\mu$ and $\tau\to\mu\gamma$ turn out to be very small in most regions of parameter space \cite{Heeck:2014qea} due to the small lepton masses involved, the $Z'$ contributions to $\tau\to 3\mu$ can be sizable~\cite{Dutta:1994dx} and restrict $\theta_R^2/v_\Phi^4$. The branching ratio is given by
\begin{align}
{\rm{Br}}\left[ {\tau  \to 3\mu } \right] \simeq \frac{{m_\tau ^5}}{{512{\pi ^3}{\Gamma _\tau }}}\frac{{{{g'}^4}}}{{m_{Z'}^4}}{\sin ^2}\left( {2{\theta _R}} \right) ,
\end{align}
which has to be compared to the current upper limit of $2.1\times 10^{-8}$ at $90\%$~C.L.~\cite{Aubert:2009ag} obtained by Belle. A combination with data from BaBar~\cite{Lees:2010ez} gives an even stronger limit of $1.2\times 10^{-8}$ at $90\%$~C.L.~\cite{Amhis:2014hma}, to be used in the following.
For small $\theta_R$, the branching ratio for $\tau\to\mu\gamma$ is proportional to the same combination $\theta_R^2/v_\Phi^4$, but highly suppressed by $2 \alpha/\pi$, and hence not as restrictive.

\subsection{\BKs and $B\to K\mu^+\mu^-/B\to K e^+ e^-$}

Both \BKs and $R(K)$ are sensitive to the Wilson coefficients $C^{(\prime)\mu\mu}_9$ and $C^{(\prime)\mu\mu}_{10}$.\footnote{For conventions see Refs.~\cite{Hiller:2014yaa,Altmannshofer:2014rta}.}
While in our model the contribution to $C_{10}$ is suppressed by $\sin(2\theta_{R})$ (or even $\sin(2\theta_{L})$), the Wilson coefficients $C_9^{\mu\mu}$ and $C_9^{\prime\mu\mu}$ with muons are generated (as well as $C_9^{\tau\tau}$ and the $\theta_R$ suppressed $C_9^{\mu\tau}$). $C_9^{\prime e e}$ is not affected, which naturally generates violations of lepton flavour universality in $B\to K\mu^+\mu^-/B\to K e^+ e^-$. We find
\begin{equation}
{C^{(\prime)\mu\mu}_9} \simeq \frac{{ {{g'}^2}}}{{\sqrt 2 m_{Z'}^2}}\frac{\pi }{\alpha }\frac{1}{{{G_F}{V_{tb}}V_{ts}^*}}\Gamma_{23}^{d L(R)}\,,
\label{eq:C9}
\end{equation}
where we set $\cos(2\theta_{R})=1$. As already noted in Ref.~\cite{Descotes-Genon:2013wba,Descotes-Genon:2013zva} $C^{\mu\mu}_9<0$ and $C^{\prime\mu\mu}_9=0$ gives a good fit to data. Using the global fit of Ref.~\cite{Altmannshofer:2014rta} we see that at ($1\,\sigma$) $2\,\sigma$ level 
\begin{equation}
 -0.5\, (-0.8) \geq \real C_9^{\mu\mu}	\geq(-1.6)\, -2.0 \,.
\label{eq:C9fit}
\end{equation}
Interestingly, the regions for $C_9^{\mu\mu}$ required by $R(K)$ and \BKs lie approximately in the same region. Furthermore, a good fit to the current data does not even require $C^{\prime \mu\mu}_9$~\cite{Altmannshofer:2014rta}, so we neglect it in the following for simplicity. This can be achieved in the limit $m_{D} \gg m_{Q}$, resulting in $\Gamma^{d L} \gg \Gamma^{d R}$. We will also assume our $C_9^{\mu\mu}$ to be real for simplicity. Note that our model predicts the decay $B\to K \mu \tau$ (recently discussed in Refs.~\cite{Glashow:2014iga}) to be suppressed by $\theta_R^2$ compared to $B\to K\mu\mu$, while $B\to K\mu e$ and $B\to K\tau e$ are forbidden. 

\subsection{$B_s$--$\overline{B}_s$ mixing}

The interactions of $Z'$ and $\Phi$ relevant for $B\to K \mu^+\mu^-$ also contribute to $B_s$--$\overline{B}_s$ mixing~\cite{Altmannshofer:2014cfa}. For $m_{D} \gg m_{Q}$, we get
\begin{align}
\frac{{{M_{12}}}}{{M_{12}^\text{SM}}} \simeq 1 + \frac{{{{\left( {\Gamma _{23}^{dL}} \right)}^2}\left( {1 + \frac{1}{{16{\pi ^2}}} \frac{{{{g'}^2}{m_Q^2}}}{{m_{Z'}^2}}} \right)}}{{\frac{{g_2^4}}{{64{\pi ^2}}}\frac{{m_{Z'}^2}}{{m_W^2{{g'}^2}}}{{\left( {V_{ts}^*V_{tb}^{}} \right)}^2}{S_0}}} \,.
\end{align}
We require the NP contribution to be less than $15\%$ in order to satisfy the experimental bounds~\cite{Altmannshofer:2014cfa}. Due to the dominance of the vector-quark $Q$ we can express $\Gamma_{23}^{dL}$ directly in terms of $C_9^{\mu\mu}$ from \eq{eq:C9} and find the \emph{upper} bounds
\begin{align}
{m_{Z'}}/{g'} < {3.2 \TeV}/{|C_9^{\mu\mu}|},\;\; m_Q < {41 \TeV}/{|C_9^{\mu\mu}|} \,.
\label{eq:upperVEV}
\end{align}
Combining \eq{eq:upperVEV} with \eq{eq:C9fit} then gives an upper bound of $m_{Z'}/g' < 4\TeV$ ($6.5\TeV$) at $1\,\sigma$ ($2\,\sigma$).

\subsection{Neutrino trident production}

\begin{figure}[t]
\includegraphics[width=0.46\textwidth]{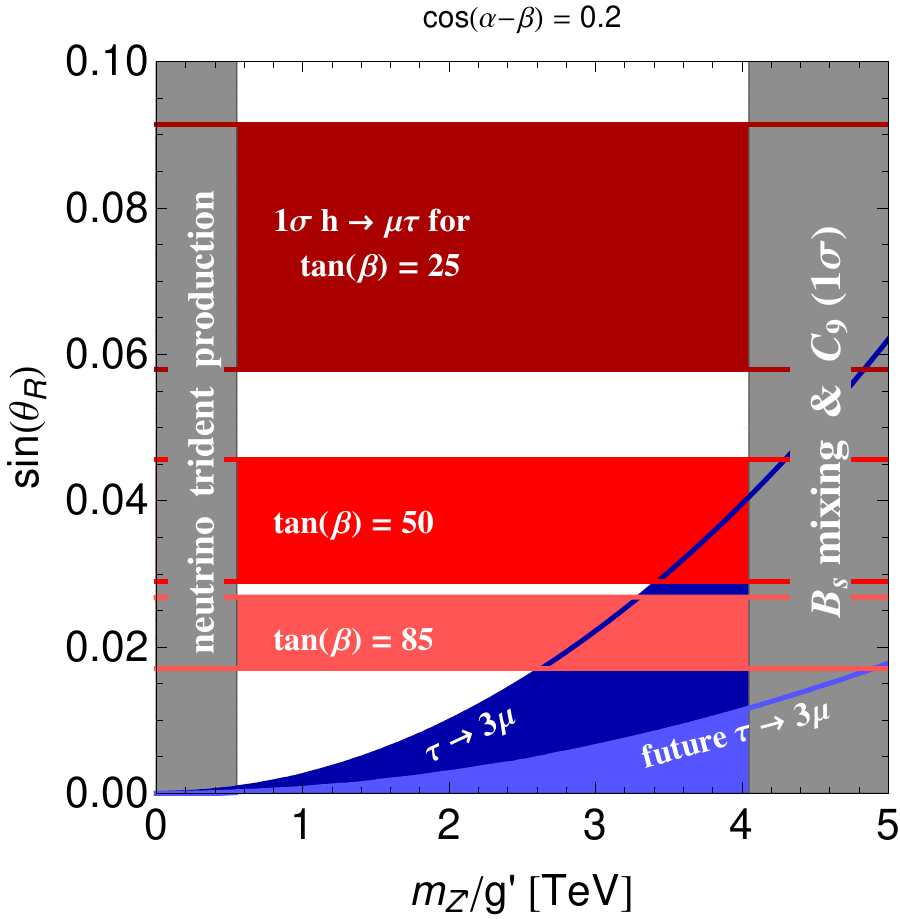}
\caption{Allowed regions in the $m_{Z'}/g'$--$\sin (\theta_R)$ plane: the horizontal stripes correspond to $h\to\mu\tau$ ($1\sigma$) for $\tan\beta=85,\,50,\,25$ and $\cos (\alpha-\beta)=0.2$, (light) blue stands for (future) $\tau\to 3\mu$ limits at $90\%$~C.L. The gray regions are excluded by NTP or $B_s$--$\overline{B}_s$ mixing in combination with the $1\,\sigma$ range for $C_9$ (see \eq{eq:upperVEV}). }
\label{fig:vevplot}
\end{figure}

The most stringent bound on flavour-diagonal $Z'$ couplings to muons arises from neutrino trident production (NTP) $\nu_\mu N \to \nu_\mu N \mu^+\mu^-$~\cite{Altmannshofer:2014cfa, Altmannshofer:2014pba}:
\begin{equation}
	\frac{\sigma_{\rm NTP}}{\sigma_{\rm NTP}^{\rm SM}}\simeq \frac{1+\left(1+4s_W^2+8\frac{g'^2}{M_{Z^\prime}^2}\frac{m_W^2}{g_2^2}\right)^2}{1+\left(1+4s_W^2\right)^2}\,.
\end{equation}
Seeing as our region of interest is in the small $\theta_{R}$ regime, the NTP bound is basically independent of the angle $\theta_R$. Taking only the CCFR data~\cite{Mishra:1991bv}, we get roughly $m_{Z'}/g' \gtrsim 550 \GeV$ at $95\%$~C.L. Compared to $\tau\to\mu\mu\mu$ the trident neutrino bound only dominates for very small values of $\theta_R$, roughly when $\theta_R \lesssim 10^{-3}$ (see Fig.~\ref{fig:vevplot} (right)).

For $m_{Z'} > m_Z$, the LHC constraints from the process $pp \to \mu\mu Z' \to 4\mu$ (or $3\mu$ plus missing energy)~\cite{Ma:2001md} are currently weaker than NTP~\cite{Altmannshofer:2014cfa}, but will become competitive with higher luminosities~\cite{Harigaya:2013twa,Bell:2014tta,delAguila:2014soa}.

\subsection{Phenomenological analysis}
\label{pheno}

Concerning the phenomenological consequences of of our model, let us first consider the implications of $h\to\mu\tau$. In the left plot of Fig.~\ref{fig:HiggsPlot} we show the regions in the $\cos(\alpha-\beta)$--$\sin(\theta_R)$ plane which can explain $h\to\mu\tau$ at the $1\,\sigma$ and $2\,\sigma$ level for different values of $\tan\beta$. Measurements of the $h$ couplings to vector bosons require $|\cos(\alpha-\beta)| \lesssim 0.4$~\cite{Dumont:2014wha,Dumont:2014kna} while the Higgs effects in $\tau\to3\mu$ and $\tau\to\mu\gamma$ are typically negligible~\cite{Heeck:2014qea}. As a side effect, the $h\to\mu\tau$ rate also implies a change in the $h\to\tau\tau$ rate, although this is negligible in regions with small $\theta_R$. In addition we show the regions compatible with $\tau\to3\mu$ for various values of $m_{Z'}/g'$. Note that $g'\lesssim 0.3$ in order to avoid a Landau pole below the Planck scale. In summary, small values of $\theta_R$ can explain the CMS $h\to\mu\tau$ excess for moderate to large values of $\tan\beta$ for $\cos(\alpha-\beta)\simeq 0.1$. 

In the right plot of Fig.~\ref{fig:HiggsPlot} we examine which regions in parameter space can account for \BKs taking into account the constraints from $B_s$--$\bar B_s$ mixing. Since we focus on the limit $M_D\to\infty$ (i.e.~$C_9^{\prime}\to0$) we find that unless $\Gamma^{dL}_{23}$ is rather large, \BKs can be explained without violating bounds from $B_s$--$\bar B_s$. Only a very small $\Gamma^{dL}_{23}$ independent region is excluded by NTP. In addition, bounds from $\tau\to3\mu$ depending on $\sin(\theta_R)$ can be obtained.

Concerning $\tau\to 3\mu$, future sensitivities down to ${\rm{Br}}\left[ {\tau  \to 3\mu } \right]\simeq 10^{-9}$ seem feasible~\cite{Aushev:2010bq} and will cut deep into our parameter space (see Fig.~\ref{fig:vevplot}). 
Using the $1\, \sigma$ limits on $h\to\mu\tau$ to fix $\theta_R$ and $B_s$ mixing with $C_9$ to fix $m_{Z'}/g'$ -- as well as the LHC limit $|\cos(\alpha-\beta)|<0.4$ -- we can obtain a \emph{lower} limit on the rate $\tau\to 3\mu$
\begin{align}
{\rm{Br}}\left[ {\tau  \to 3\mu } \right] \gtrsim 3.8 \times 10^{-8} \left(10/\tan \beta\right)^2 ,
\end{align}
which implies $\tan\beta \gtrsim 18$ with current data~\cite{Amhis:2014hma} and $\tan\beta \gtrsim 61$ if branching ratios down to $10^{-9}$ can be probed in the future. This is the main prediction of our simultaneous explanation of $h\to\mu\tau$, \BKs and $R(K)$.

Finally, we remark that a $Z$--$Z'$ mixing angle $\theta_{Z Z'}$~\cite{Langacker:2008yv} is induced by the VEV of $\Psi_1$~\cite{Heeck:2014qea}
\begin{align}
|g' \theta_{Z Z'} |\simeq \frac{g_1 v^2 \cos^2\beta}{m_{Z'}^2/g'^2} \simeq 10^{-4} \left(\frac{20}{\tan\beta}\right)^2 \left(\frac{\TeV}{m_{Z'}/g'}\right)^2 ,
\end{align}
which leads to small shifts in the vector couplings of $Z$ to muons and taus
\begin{align}
g_V^Z (\mu\mu, \tau\tau) \simeq -1/2 + 2 s_W^2 \pm g' \theta_{Z Z'}/(g/c_W) \,,
\end{align}
and thus ultimately to lepton non-universality~\cite{Heeck:2011wj}. For the values of interest to our study (see Fig.~\ref{fig:vevplot}), and in the limit $m_Z \ll m_{Z'}$, the shift is automatically small enough to satisfy experimental bounds and leads to tiny branching ratios $Z\to \mu\tau$ below $10^{-8}$ (for $\theta_R<0.1$).
Note that the couplings to electrons and quarks remain unaffected. For $m_{Z'} \gg m_Z$, the $\rho$ parameter is enhanced by~\cite{Langacker:2008yv}
\begin{align}
\rho -1 \simeq 1.2\times 10^{-4} \left( \frac{\theta_{Z Z'}}{10^{-3}}\right)^2 \left(\frac{m_{Z'}}{\TeV}\right)^2 ,
\end{align}
and is therefore compatible with electroweak precision data ($\rho -1 < 9\times 10^{-4}$ at $2\sigma$~\cite{Agashe:2014kda}) for the parameter space studied in this letter.

\section{Conclusions}
\label{conclusion}

In this letter we showed for the first time that all three LHC anomalies in the flavour sector can be explained within a single well-motivated model: A 2HDM with a gauged $L_\mu-L_\tau$ symmetry and effective $Z^\prime\bar s b$ couplings induced by heavy vector-like quarks. Except for the $\tau$--$\mu$ couplings, the Higgs sector resembles the one of a 2HDM of type-I. Therefore, the constraints from $h$ decays or LHC searches for $A^0\to\tau^+\tau^-$ are rather weak and $h\to\mu\tau$ can be easily explained in a wide parameter space. The model can also account for the deviations from the SM in \BKs and naturally leads to the right amount of lepton-flavour-universality violating effects in $R(K)$. Due to the small values of the $\tau$--$\mu$ mixing angle $\theta_R$, sufficient to account for $h\to\mu\tau$, the $Z'$ contributions to $\tau\to3\mu$ are not in conflict with present bounds for large $\tan\beta$ in wide rages of parameter space. Interestingly, \BKs and $R(K)$ combined with $B_s$--$\overline{B}_s$ put a upper limit on $m_{Z'}/g'$ resulting in a lower limit on $\tau\to3\mu$ if ${\rm Br}[h\to\mu\tau]\neq0$: for lower values of $\tan\beta$ the current experimental bounds are reached and future sensitivities will allow for a more detailed exploration of the allowed parameter space. The possible range for the $L_\mu-L_\tau$ breaking scale further implies the masses of the $Z'$ and the right-handed neutrinos to be at the TeV scale, potentially testable at the LHC with interesting additional consequences for LFV observables.

\acknowledgments{A.~Crivellin is supported by a Marie Curie Intra-European Fellowship of the European Community's 7th Framework Programme under contract number PIEF-GA-2012-326948. G.~D'Ambrosio acknowledges the partial support my MIUR under the project number 2010YJ2NYW. The work of J.~Heeck is funded in part by IISN and by Belgian Science Policy (IAP VII/37). We thank Gian Giudice for useful discussions. We are grateful to Wolfgang Altmannshofer and David Straub for useful discussions and additional information concerning the model-independent fit to $C^{\mu\mu}_9$ and $C^{\prime\mu\mu}_9$.}
\\\emph{Note added}:
during the publication process of this letter, CMS has released its final analysis of the $h\to\mu\tau$ search as a preprint~\cite{Khachatryan:2015kon}, resulting in slightly changed values -- ${\rm Br} [h\to\mu\tau] = \left( 0.84_{-0.37}^{+0.39} \right)\%$ -- which have however only a small impact on our study.

\bibliography{BIB}

\begin{thebibliography}{63}
\expandafter\ifx\csname natexlab\endcsname\relax\def\natexlab#1{#1}\fi
\expandafter\ifx\csname bibnamefont\endcsname\relax
  \def\bibnamefont#1{#1}\fi
\expandafter\ifx\csname bibfnamefont\endcsname\relax
  \def\bibfnamefont#1{#1}\fi
\expandafter\ifx\csname citenamefont\endcsname\relax
  \def\citenamefont#1{#1}\fi
\expandafter\ifx\csname url\endcsname\relax
  \def\url#1{\texttt{#1}}\fi
\expandafter\ifx\csname urlprefix\endcsname\relax\def\urlprefix{URL }\fi
\providecommand{\bibinfo}[2]{#2}
\providecommand{\eprint}[2][]{\url{#2}}

\bibitem[{\citenamefont{Aad et~al.}(2012)}]{Aad:2012tfa}
\bibinfo{author}{\bibfnamefont{G.}~\bibnamefont{Aad}} \bibnamefont{et~al.}
  (\bibinfo{collaboration}{ATLAS Collaboration}), \bibinfo{journal}{Phys.Lett.}
  \textbf{\bibinfo{volume}{B716}}, \bibinfo{pages}{1} (\bibinfo{year}{2012}),
  \eprint{1207.7214}.

\bibitem[{\citenamefont{Chatrchyan et~al.}(2012)}]{Chatrchyan:2012ufa}
\bibinfo{author}{\bibfnamefont{S.}~\bibnamefont{Chatrchyan}}
  \bibnamefont{et~al.} (\bibinfo{collaboration}{CMS Collaboration}),
  \bibinfo{journal}{Phys.Lett.} \textbf{\bibinfo{volume}{B716}},
  \bibinfo{pages}{30} (\bibinfo{year}{2012}), \eprint{1207.7235}.

\bibitem[{\citenamefont{Egede et~al.}(2008)\citenamefont{Egede, Hurth, Matias,
  Ramon, and Reece}}]{Egede:2008uy}
\bibinfo{author}{\bibfnamefont{U.}~\bibnamefont{Egede}},
  \bibinfo{author}{\bibfnamefont{T.}~\bibnamefont{Hurth}},
  \bibinfo{author}{\bibfnamefont{J.}~\bibnamefont{Matias}},
  \bibinfo{author}{\bibfnamefont{M.}~\bibnamefont{Ramon}}, \bibnamefont{and}
  \bibinfo{author}{\bibfnamefont{W.}~\bibnamefont{Reece}},
  \bibinfo{journal}{JHEP} \textbf{\bibinfo{volume}{0811}}, \bibinfo{pages}{032}
  (\bibinfo{year}{2008}), \eprint{0807.2589}.

\bibitem[{\citenamefont{Altmannshofer et~al.}(2009)\citenamefont{Altmannshofer,
  Ball, Bharucha, Buras, Straub et~al.}}]{Altmannshofer:2008dz}
\bibinfo{author}{\bibfnamefont{W.}~\bibnamefont{Altmannshofer}},
  \bibinfo{author}{\bibfnamefont{P.}~\bibnamefont{Ball}},
  \bibinfo{author}{\bibfnamefont{A.}~\bibnamefont{Bharucha}},
  \bibinfo{author}{\bibfnamefont{A.~J.} \bibnamefont{Buras}},
  \bibinfo{author}{\bibfnamefont{D.~M.} \bibnamefont{Straub}},
  \bibnamefont{et~al.}, \bibinfo{journal}{JHEP}
  \textbf{\bibinfo{volume}{0901}}, \bibinfo{pages}{019} (\bibinfo{year}{2009}),
  \eprint{0811.1214}.

\bibitem[{\citenamefont{Descotes-Genon
  et~al.}(2013{\natexlab{a}})\citenamefont{Descotes-Genon, Hurth, Matias, and
  Virto}}]{Descotes-Genon:2013vna}
\bibinfo{author}{\bibfnamefont{S.}~\bibnamefont{Descotes-Genon}},
  \bibinfo{author}{\bibfnamefont{T.}~\bibnamefont{Hurth}},
  \bibinfo{author}{\bibfnamefont{J.}~\bibnamefont{Matias}}, \bibnamefont{and}
  \bibinfo{author}{\bibfnamefont{J.}~\bibnamefont{Virto}},
  \bibinfo{journal}{JHEP} \textbf{\bibinfo{volume}{1305}}, \bibinfo{pages}{137}
  (\bibinfo{year}{2013}{\natexlab{a}}), \eprint{1303.5794}.

\bibitem[{\citenamefont{Aaij et~al.}(2013)}]{Aaij:2013qta}
\bibinfo{author}{\bibfnamefont{R.}~\bibnamefont{Aaij}} \bibnamefont{et~al.}
  (\bibinfo{collaboration}{LHCb collaboration}),
  \bibinfo{journal}{Phys.Rev.Lett.} \textbf{\bibinfo{volume}{111}},
  \bibinfo{pages}{191801} (\bibinfo{year}{2013}), \eprint{1308.1707}.

\bibitem[{\citenamefont{Descotes-Genon
  et~al.}(2014)\citenamefont{Descotes-Genon, Hofer, Matias, and
  Virto}}]{Descotes-Genon:2014uoa}
\bibinfo{author}{\bibfnamefont{S.}~\bibnamefont{Descotes-Genon}},
  \bibinfo{author}{\bibfnamefont{L.}~\bibnamefont{Hofer}},
  \bibinfo{author}{\bibfnamefont{J.}~\bibnamefont{Matias}}, \bibnamefont{and}
  \bibinfo{author}{\bibfnamefont{J.}~\bibnamefont{Virto}},
  \bibinfo{journal}{JHEP} \textbf{\bibinfo{volume}{1412}}, \bibinfo{pages}{125}
  (\bibinfo{year}{2014}), \eprint{1407.8526}.

\bibitem[{\citenamefont{Altmannshofer and
  Straub}(2014)}]{Altmannshofer:2014rta}
\bibinfo{author}{\bibfnamefont{W.}~\bibnamefont{Altmannshofer}}
  \bibnamefont{and} \bibinfo{author}{\bibfnamefont{D.~M.} \bibnamefont{Straub}}
  (\bibinfo{year}{2014}), \eprint{1411.3161}.

\bibitem[{\citenamefont{J{\"a}ger and Martin~Camalich}(2014)}]{Jager:2014rwa}
\bibinfo{author}{\bibfnamefont{S.}~\bibnamefont{J{\"a}ger}} \bibnamefont{and}
  \bibinfo{author}{\bibfnamefont{J.}~\bibnamefont{Martin~Camalich}}
  (\bibinfo{year}{2014}), \eprint{1412.3183}.

\bibitem[{\citenamefont{Descotes-Genon
  et~al.}(2013{\natexlab{b}})\citenamefont{Descotes-Genon, Matias, and
  Virto}}]{Descotes-Genon:2013wba}
\bibinfo{author}{\bibfnamefont{S.}~\bibnamefont{Descotes-Genon}},
  \bibinfo{author}{\bibfnamefont{J.}~\bibnamefont{Matias}}, \bibnamefont{and}
  \bibinfo{author}{\bibfnamefont{J.}~\bibnamefont{Virto}},
  \bibinfo{journal}{Phys.Rev.} \textbf{\bibinfo{volume}{D88}},
  \bibinfo{pages}{074002} (\bibinfo{year}{2013}{\natexlab{b}}),
  \eprint{1307.5683}.

\bibitem[{\citenamefont{Altmannshofer and
  Straub}(2013)}]{Altmannshofer:2013foa}
\bibinfo{author}{\bibfnamefont{W.}~\bibnamefont{Altmannshofer}}
  \bibnamefont{and} \bibinfo{author}{\bibfnamefont{D.~M.}
  \bibnamefont{Straub}}, \bibinfo{journal}{Eur.Phys.J.}
  \textbf{\bibinfo{volume}{C73}}, \bibinfo{pages}{2646} (\bibinfo{year}{2013}),
  \eprint{1308.1501}.

\bibitem[{\citenamefont{Horgan et~al.}(2014)\citenamefont{Horgan, Liu, Meinel,
  and Wingate}}]{Horgan:2013pva}
\bibinfo{author}{\bibfnamefont{R.~R.} \bibnamefont{Horgan}},
  \bibinfo{author}{\bibfnamefont{Z.}~\bibnamefont{Liu}},
  \bibinfo{author}{\bibfnamefont{S.}~\bibnamefont{Meinel}}, \bibnamefont{and}
  \bibinfo{author}{\bibfnamefont{M.}~\bibnamefont{Wingate}},
  \bibinfo{journal}{Phys.Rev.Lett.} \textbf{\bibinfo{volume}{112}},
  \bibinfo{pages}{212003} (\bibinfo{year}{2014}), \eprint{1310.3887}.

\bibitem[{\citenamefont{Gauld et~al.}(2014)\citenamefont{Gauld, Goertz, and
  Haisch}}]{Gauld:2013qba}
\bibinfo{author}{\bibfnamefont{R.}~\bibnamefont{Gauld}},
  \bibinfo{author}{\bibfnamefont{F.}~\bibnamefont{Goertz}}, \bibnamefont{and}
  \bibinfo{author}{\bibfnamefont{U.}~\bibnamefont{Haisch}},
  \bibinfo{journal}{Phys.Rev.} \textbf{\bibinfo{volume}{D89}},
  \bibinfo{pages}{015005} (\bibinfo{year}{2014}), \eprint{1308.1959}.

\bibitem[{\citenamefont{Buras et~al.}(2014)\citenamefont{Buras, De~Fazio, and
  Girrbach}}]{Buras:2013dea}
\bibinfo{author}{\bibfnamefont{A.~J.} \bibnamefont{Buras}},
  \bibinfo{author}{\bibfnamefont{F.}~\bibnamefont{De~Fazio}}, \bibnamefont{and}
  \bibinfo{author}{\bibfnamefont{J.}~\bibnamefont{Girrbach}},
  \bibinfo{journal}{JHEP} \textbf{\bibinfo{volume}{1402}}, \bibinfo{pages}{112}
  (\bibinfo{year}{2014}), \eprint{1311.6729}.

\bibitem[{\citenamefont{Altmannshofer
  et~al.}(2014{\natexlab{a}})\citenamefont{Altmannshofer, Gori, Pospelov, and
  Yavin}}]{Altmannshofer:2014cfa}
\bibinfo{author}{\bibfnamefont{W.}~\bibnamefont{Altmannshofer}},
  \bibinfo{author}{\bibfnamefont{S.}~\bibnamefont{Gori}},
  \bibinfo{author}{\bibfnamefont{M.}~\bibnamefont{Pospelov}}, \bibnamefont{and}
  \bibinfo{author}{\bibfnamefont{I.}~\bibnamefont{Yavin}},
  \bibinfo{journal}{Phys.Rev.} \textbf{\bibinfo{volume}{D89}},
  \bibinfo{pages}{095033} (\bibinfo{year}{2014}{\natexlab{a}}),
  \eprint{1403.1269}.

\bibitem[{\citenamefont{Aaij et~al.}(2014)}]{Aaij:2014ora}
\bibinfo{author}{\bibfnamefont{R.}~\bibnamefont{Aaij}} \bibnamefont{et~al.}
  (\bibinfo{collaboration}{LHCb collaboration}),
  \bibinfo{journal}{Phys.Rev.Lett.} \textbf{\bibinfo{volume}{113}},
  \bibinfo{pages}{151601} (\bibinfo{year}{2014}), \eprint{1406.6482}.

\bibitem[{\citenamefont{Bobeth et~al.}(2007)\citenamefont{Bobeth, Hiller, and
  Piranishvili}}]{Bobeth:2007dw}
\bibinfo{author}{\bibfnamefont{C.}~\bibnamefont{Bobeth}},
  \bibinfo{author}{\bibfnamefont{G.}~\bibnamefont{Hiller}}, \bibnamefont{and}
  \bibinfo{author}{\bibfnamefont{G.}~\bibnamefont{Piranishvili}},
  \bibinfo{journal}{JHEP} \textbf{\bibinfo{volume}{0712}}, \bibinfo{pages}{040}
  (\bibinfo{year}{2007}), \eprint{0709.4174}.

\bibitem[{\citenamefont{Alonso et~al.}(2014)\citenamefont{Alonso, Grinstein,
  and Martin~Camalich}}]{Alonso:2014csa}
\bibinfo{author}{\bibfnamefont{R.}~\bibnamefont{Alonso}},
  \bibinfo{author}{\bibfnamefont{B.}~\bibnamefont{Grinstein}},
  \bibnamefont{and}
  \bibinfo{author}{\bibfnamefont{J.}~\bibnamefont{Martin~Camalich}},
  \bibinfo{journal}{Phys.Rev.Lett.} \textbf{\bibinfo{volume}{113}},
  \bibinfo{pages}{241802} (\bibinfo{year}{2014}), \eprint{1407.7044}.

\bibitem[{\citenamefont{Hiller and Schmaltz}(2014)}]{Hiller:2014yaa}
\bibinfo{author}{\bibfnamefont{G.}~\bibnamefont{Hiller}} \bibnamefont{and}
  \bibinfo{author}{\bibfnamefont{M.}~\bibnamefont{Schmaltz}},
  \bibinfo{journal}{Phys.Rev.} \textbf{\bibinfo{volume}{D90}},
  \bibinfo{pages}{054014} (\bibinfo{year}{2014}), \eprint{1408.1627}.

\bibitem[{\citenamefont{Ghosh et~al.}(2014)\citenamefont{Ghosh, Nardecchia, and
  Renner}}]{Ghosh:2014awa}
\bibinfo{author}{\bibfnamefont{D.}~\bibnamefont{Ghosh}},
  \bibinfo{author}{\bibfnamefont{M.}~\bibnamefont{Nardecchia}},
  \bibnamefont{and} \bibinfo{author}{\bibfnamefont{S.}~\bibnamefont{Renner}},
  \bibinfo{journal}{JHEP} \textbf{\bibinfo{volume}{1412}}, \bibinfo{pages}{131}
  (\bibinfo{year}{2014}), \eprint{1408.4097}.

\bibitem[{\citenamefont{Hurth et~al.}(2014)\citenamefont{Hurth, Mahmoudi, and
  Neshatpour}}]{Hurth:2014vma}
\bibinfo{author}{\bibfnamefont{T.}~\bibnamefont{Hurth}},
  \bibinfo{author}{\bibfnamefont{F.}~\bibnamefont{Mahmoudi}}, \bibnamefont{and}
  \bibinfo{author}{\bibfnamefont{S.}~\bibnamefont{Neshatpour}},
  \bibinfo{journal}{JHEP} \textbf{\bibinfo{volume}{1412}}, \bibinfo{pages}{053}
  (\bibinfo{year}{2014}), \eprint{1410.4545}.

\bibitem[{\citenamefont{CMS}(2014)}]{CMS:2014hha}
\bibinfo{author}{\bibnamefont{CMS}} (\bibinfo{collaboration}{CMS
  Collaboration}) (\bibinfo{year}{2014}), \bibinfo{note}{{CMS-PAS-HIG-14-005}}.

\bibitem[{\citenamefont{Harnik et~al.}(2013)\citenamefont{Harnik, Kopp, and
  Zupan}}]{Harnik:2012pb}
\bibinfo{author}{\bibfnamefont{R.}~\bibnamefont{Harnik}},
  \bibinfo{author}{\bibfnamefont{J.}~\bibnamefont{Kopp}}, \bibnamefont{and}
  \bibinfo{author}{\bibfnamefont{J.}~\bibnamefont{Zupan}},
  \bibinfo{journal}{JHEP} \textbf{\bibinfo{volume}{1303}}, \bibinfo{pages}{026}
  (\bibinfo{year}{2013}), \eprint{1209.1397}.

\bibitem[{\citenamefont{Blankenburg et~al.}(2012)\citenamefont{Blankenburg,
  Ellis, and Isidori}}]{Blankenburg:2012ex}
\bibinfo{author}{\bibfnamefont{G.}~\bibnamefont{Blankenburg}},
  \bibinfo{author}{\bibfnamefont{J.}~\bibnamefont{Ellis}}, \bibnamefont{and}
  \bibinfo{author}{\bibfnamefont{G.}~\bibnamefont{Isidori}},
  \bibinfo{journal}{Phys.Lett.} \textbf{\bibinfo{volume}{B712}},
  \bibinfo{pages}{386} (\bibinfo{year}{2012}), \eprint{1202.5704}.

\bibitem[{\citenamefont{Davidson and Verdier}(2012)}]{Davidson:2012ds}
\bibinfo{author}{\bibfnamefont{S.}~\bibnamefont{Davidson}} \bibnamefont{and}
  \bibinfo{author}{\bibfnamefont{P.}~\bibnamefont{Verdier}},
  \bibinfo{journal}{Phys.Rev.} \textbf{\bibinfo{volume}{D86}},
  \bibinfo{pages}{111701} (\bibinfo{year}{2012}), \eprint{1211.1248}.

\bibitem[{\citenamefont{Arhrib et~al.}(2013{\natexlab{a}})\citenamefont{Arhrib,
  Cheng, and Kong}}]{Arhrib:2012ax}
\bibinfo{author}{\bibfnamefont{A.}~\bibnamefont{Arhrib}},
  \bibinfo{author}{\bibfnamefont{Y.}~\bibnamefont{Cheng}}, \bibnamefont{and}
  \bibinfo{author}{\bibfnamefont{O.~C.} \bibnamefont{Kong}},
  \bibinfo{journal}{Phys.Rev.} \textbf{\bibinfo{volume}{D87}},
  \bibinfo{pages}{015025} (\bibinfo{year}{2013}{\natexlab{a}}),
  \eprint{1210.8241}.

\bibitem[{\citenamefont{Arhrib et~al.}(2013{\natexlab{b}})\citenamefont{Arhrib,
  Cheng, and Kong}}]{Arhrib:2012mg}
\bibinfo{author}{\bibfnamefont{A.}~\bibnamefont{Arhrib}},
  \bibinfo{author}{\bibfnamefont{Y.}~\bibnamefont{Cheng}}, \bibnamefont{and}
  \bibinfo{author}{\bibfnamefont{O.~C.} \bibnamefont{Kong}},
  \bibinfo{journal}{Europhys.Lett.} \textbf{\bibinfo{volume}{101}},
  \bibinfo{pages}{31003} (\bibinfo{year}{2013}{\natexlab{b}}),
  \eprint{1208.4669}.

\bibitem[{\citenamefont{Kopp and Nardecchia}(2014)}]{Kopp:2014rva}
\bibinfo{author}{\bibfnamefont{J.}~\bibnamefont{Kopp}} \bibnamefont{and}
  \bibinfo{author}{\bibfnamefont{M.}~\bibnamefont{Nardecchia}},
  \bibinfo{journal}{JHEP} \textbf{\bibinfo{volume}{1410}}, \bibinfo{pages}{156}
  (\bibinfo{year}{2014}), \eprint{1406.5303}.

\bibitem[{\citenamefont{Falkowski et~al.}(2014)\citenamefont{Falkowski, Straub,
  and Vicente}}]{Falkowski:2013jya}
\bibinfo{author}{\bibfnamefont{A.}~\bibnamefont{Falkowski}},
  \bibinfo{author}{\bibfnamefont{D.~M.} \bibnamefont{Straub}},
  \bibnamefont{and} \bibinfo{author}{\bibfnamefont{A.}~\bibnamefont{Vicente}},
  \bibinfo{journal}{JHEP} \textbf{\bibinfo{volume}{1405}}, \bibinfo{pages}{092}
  (\bibinfo{year}{2014}), \eprint{1312.5329}.

\bibitem[{\citenamefont{Dery et~al.}(2014)\citenamefont{Dery, Efrati, Nir,
  Soreq, and Susič}}]{Dery:2014kxa}
\bibinfo{author}{\bibfnamefont{A.}~\bibnamefont{Dery}},
  \bibinfo{author}{\bibfnamefont{A.}~\bibnamefont{Efrati}},
  \bibinfo{author}{\bibfnamefont{Y.}~\bibnamefont{Nir}},
  \bibinfo{author}{\bibfnamefont{Y.}~\bibnamefont{Soreq}}, \bibnamefont{and}
  \bibinfo{author}{\bibfnamefont{V.}~\bibnamefont{Susič}},
  \bibinfo{journal}{Phys.Rev.} \textbf{\bibinfo{volume}{D90}},
  \bibinfo{pages}{115022} (\bibinfo{year}{2014}), \eprint{1408.1371}.

\bibitem[{\citenamefont{Campos et~al.}(2014)\citenamefont{Campos,
  Hern{\'a}ndez, P{\"a}s, and Schumacher}}]{Campos:2014zaa}
\bibinfo{author}{\bibfnamefont{M.~D.} \bibnamefont{Campos}},
  \bibinfo{author}{\bibfnamefont{A.~E.~C.} \bibnamefont{Hern{\'a}ndez}},
  \bibinfo{author}{\bibfnamefont{H.}~\bibnamefont{P{\"a}s}}, \bibnamefont{and}
  \bibinfo{author}{\bibfnamefont{E.}~\bibnamefont{Schumacher}}
  (\bibinfo{year}{2014}), \eprint{1408.1652}.

\bibitem[{\citenamefont{Celis et~al.}(2014)\citenamefont{Celis, Cirigliano, and
  Passemar}}]{Celis:2014roa}
\bibinfo{author}{\bibfnamefont{A.}~\bibnamefont{Celis}},
  \bibinfo{author}{\bibfnamefont{V.}~\bibnamefont{Cirigliano}},
  \bibnamefont{and} \bibinfo{author}{\bibfnamefont{E.}~\bibnamefont{Passemar}}
  (\bibinfo{year}{2014}), \eprint{1409.4439}.

\bibitem[{\citenamefont{Aristizabal~Sierra and Vicente}(2014)}]{Sierra:2014nqa}
\bibinfo{author}{\bibfnamefont{D.}~\bibnamefont{Aristizabal~Sierra}}
  \bibnamefont{and} \bibinfo{author}{\bibfnamefont{A.}~\bibnamefont{Vicente}},
  \bibinfo{journal}{Phys.Rev.} \textbf{\bibinfo{volume}{D90}},
  \bibinfo{pages}{115004} (\bibinfo{year}{2014}), \eprint{1409.7690}.

\bibitem[{\citenamefont{Lee and Tandean}(2014)}]{Lee:2014rba}
\bibinfo{author}{\bibfnamefont{C.-J.} \bibnamefont{Lee}} \bibnamefont{and}
  \bibinfo{author}{\bibfnamefont{J.}~\bibnamefont{Tandean}}
  (\bibinfo{year}{2014}), \eprint{1410.6803}.

\bibitem[{\citenamefont{Heeck et~al.}(2014)\citenamefont{Heeck, Holthausen,
  Rodejohann, and Shimizu}}]{Heeck:2014qea}
\bibinfo{author}{\bibfnamefont{J.}~\bibnamefont{Heeck}},
  \bibinfo{author}{\bibfnamefont{M.}~\bibnamefont{Holthausen}},
  \bibinfo{author}{\bibfnamefont{W.}~\bibnamefont{Rodejohann}},
  \bibnamefont{and} \bibinfo{author}{\bibfnamefont{Y.}~\bibnamefont{Shimizu}}
  (\bibinfo{year}{2014}), \eprint{1412.3671}.

\bibitem[{\citenamefont{He et~al.}(1991{\natexlab{a}})\citenamefont{He, Joshi,
  Lew, and Volkas}}]{He:1990pn}
\bibinfo{author}{\bibfnamefont{X.}~\bibnamefont{He}},
  \bibinfo{author}{\bibfnamefont{G.~C.} \bibnamefont{Joshi}},
  \bibinfo{author}{\bibfnamefont{H.}~\bibnamefont{Lew}}, \bibnamefont{and}
  \bibinfo{author}{\bibfnamefont{R.}~\bibnamefont{Volkas}},
  \bibinfo{journal}{Phys.Rev.} \textbf{\bibinfo{volume}{D43}},
  \bibinfo{pages}{22} (\bibinfo{year}{1991}{\natexlab{a}}).

\bibitem[{\citenamefont{Foot}(1991)}]{Foot:1990mn}
\bibinfo{author}{\bibfnamefont{R.}~\bibnamefont{Foot}},
  \bibinfo{journal}{Mod.Phys.Lett.} \textbf{\bibinfo{volume}{A6}},
  \bibinfo{pages}{527} (\bibinfo{year}{1991}).

\bibitem[{\citenamefont{He et~al.}(1991{\natexlab{b}})\citenamefont{He, Joshi,
  Lew, and Volkas}}]{He:1991qd}
\bibinfo{author}{\bibfnamefont{X.-G.} \bibnamefont{He}},
  \bibinfo{author}{\bibfnamefont{G.~C.} \bibnamefont{Joshi}},
  \bibinfo{author}{\bibfnamefont{H.}~\bibnamefont{Lew}}, \bibnamefont{and}
  \bibinfo{author}{\bibfnamefont{R.}~\bibnamefont{Volkas}},
  \bibinfo{journal}{Phys.Rev.} \textbf{\bibinfo{volume}{D44}},
  \bibinfo{pages}{2118} (\bibinfo{year}{1991}{\natexlab{b}}).

\bibitem[{\citenamefont{Binetruy et~al.}(1997)\citenamefont{Binetruy, Lavignac,
  Petcov, and Ramond}}]{Binetruy:1996cs}
\bibinfo{author}{\bibfnamefont{P.}~\bibnamefont{Binetruy}},
  \bibinfo{author}{\bibfnamefont{S.}~\bibnamefont{Lavignac}},
  \bibinfo{author}{\bibfnamefont{S.~T.} \bibnamefont{Petcov}},
  \bibnamefont{and} \bibinfo{author}{\bibfnamefont{P.}~\bibnamefont{Ramond}},
  \bibinfo{journal}{Nucl.Phys.} \textbf{\bibinfo{volume}{B496}},
  \bibinfo{pages}{3} (\bibinfo{year}{1997}), \eprint{hep-ph/9610481}.

\bibitem[{\citenamefont{Bell and Volkas}(2001)}]{Bell:2000vh}
\bibinfo{author}{\bibfnamefont{N.~F.} \bibnamefont{Bell}} \bibnamefont{and}
  \bibinfo{author}{\bibfnamefont{R.~R.} \bibnamefont{Volkas}},
  \bibinfo{journal}{Phys.Rev.} \textbf{\bibinfo{volume}{D63}},
  \bibinfo{pages}{013006} (\bibinfo{year}{2001}), \eprint{hep-ph/0008177}.

\bibitem[{\citenamefont{Choubey and Rodejohann}(2005)}]{Choubey:2004hn}
\bibinfo{author}{\bibfnamefont{S.}~\bibnamefont{Choubey}} \bibnamefont{and}
  \bibinfo{author}{\bibfnamefont{W.}~\bibnamefont{Rodejohann}},
  \bibinfo{journal}{Eur.Phys.J.} \textbf{\bibinfo{volume}{C40}},
  \bibinfo{pages}{259} (\bibinfo{year}{2005}), \eprint{hep-ph/0411190}.

\bibitem[{\citenamefont{Dutta et~al.}(1994)\citenamefont{Dutta, Joshipura, and
  Vijaykumar}}]{Dutta:1994dx}
\bibinfo{author}{\bibfnamefont{G.}~\bibnamefont{Dutta}},
  \bibinfo{author}{\bibfnamefont{A.~S.} \bibnamefont{Joshipura}},
  \bibnamefont{and}
  \bibinfo{author}{\bibfnamefont{K.}~\bibnamefont{Vijaykumar}},
  \bibinfo{journal}{Phys.Rev.} \textbf{\bibinfo{volume}{D50}},
  \bibinfo{pages}{2109} (\bibinfo{year}{1994}), \eprint{hep-ph/9405292}.

\bibitem[{\citenamefont{Heeck and Rodejohann}(2011)}]{Heeck:2011wj}
\bibinfo{author}{\bibfnamefont{J.}~\bibnamefont{Heeck}} \bibnamefont{and}
  \bibinfo{author}{\bibfnamefont{W.}~\bibnamefont{Rodejohann}},
  \bibinfo{journal}{Phys.Rev.} \textbf{\bibinfo{volume}{D84}},
  \bibinfo{pages}{075007} (\bibinfo{year}{2011}), \eprint{1107.5238}.

\bibitem[{\citenamefont{Branco et~al.}(2012)\citenamefont{Branco, Ferreira,
  Lavoura, Rebelo, Sher et~al.}}]{Branco:2011iw}
\bibinfo{author}{\bibfnamefont{G.}~\bibnamefont{Branco}},
  \bibinfo{author}{\bibfnamefont{P.}~\bibnamefont{Ferreira}},
  \bibinfo{author}{\bibfnamefont{L.}~\bibnamefont{Lavoura}},
  \bibinfo{author}{\bibfnamefont{M.}~\bibnamefont{Rebelo}},
  \bibinfo{author}{\bibfnamefont{M.}~\bibnamefont{Sher}}, \bibnamefont{et~al.},
  \bibinfo{journal}{Phys.Rept.} \textbf{\bibinfo{volume}{516}},
  \bibinfo{pages}{1} (\bibinfo{year}{2012}), \eprint{1106.0034}.

\bibitem[{\citenamefont{Langacker}(2009)}]{Langacker:2008yv}
\bibinfo{author}{\bibfnamefont{P.}~\bibnamefont{Langacker}},
  \bibinfo{journal}{Rev.Mod.Phys.} \textbf{\bibinfo{volume}{81}},
  \bibinfo{pages}{1199} (\bibinfo{year}{2009}), \eprint{0801.1345}.

\bibitem[{\citenamefont{Buras et~al.}(2013)\citenamefont{Buras, De~Fazio, and
  Girrbach}}]{Buras:2012jb}
\bibinfo{author}{\bibfnamefont{A.~J.} \bibnamefont{Buras}},
  \bibinfo{author}{\bibfnamefont{F.}~\bibnamefont{De~Fazio}}, \bibnamefont{and}
  \bibinfo{author}{\bibfnamefont{J.}~\bibnamefont{Girrbach}},
  \bibinfo{journal}{JHEP} \textbf{\bibinfo{volume}{1302}}, \bibinfo{pages}{116}
  (\bibinfo{year}{2013}), \eprint{1211.1896}.

\bibitem[{\citenamefont{Dittmaier et~al.}(2012)\citenamefont{Dittmaier,
  Dittmaier, Mariotti, Passarino, Tanaka et~al.}}]{Dittmaier:2012vm}
\bibinfo{author}{\bibfnamefont{S.}~\bibnamefont{Dittmaier}},
  \bibinfo{author}{\bibfnamefont{S.}~\bibnamefont{Dittmaier}},
  \bibinfo{author}{\bibfnamefont{C.}~\bibnamefont{Mariotti}},
  \bibinfo{author}{\bibfnamefont{G.}~\bibnamefont{Passarino}},
  \bibinfo{author}{\bibfnamefont{R.}~\bibnamefont{Tanaka}},
  \bibnamefont{et~al.} (\bibinfo{year}{2012}), \eprint{1201.3084}.

\bibitem[{\citenamefont{Aubert et~al.}(2010)}]{Aubert:2009ag}
\bibinfo{author}{\bibfnamefont{B.}~\bibnamefont{Aubert}} \bibnamefont{et~al.}
  (\bibinfo{collaboration}{BaBar Collaboration}),
  \bibinfo{journal}{Phys.Rev.Lett.} \textbf{\bibinfo{volume}{104}},
  \bibinfo{pages}{021802} (\bibinfo{year}{2010}), \eprint{0908.2381}.

\bibitem[{\citenamefont{Lees et~al.}(2010)}]{Lees:2010ez}
\bibinfo{author}{\bibfnamefont{J.}~\bibnamefont{Lees}} \bibnamefont{et~al.}
  (\bibinfo{collaboration}{BaBar Collaboration}), \bibinfo{journal}{Phys.Rev.}
  \textbf{\bibinfo{volume}{D81}}, \bibinfo{pages}{111101}
  (\bibinfo{year}{2010}), \eprint{1002.4550}.

\bibitem[{\citenamefont{Amhis et~al.}(2014)}]{Amhis:2014hma}
\bibinfo{author}{\bibfnamefont{Y.}~\bibnamefont{Amhis}} \bibnamefont{et~al.}
  (\bibinfo{collaboration}{Heavy Flavor Averaging Group (HFAG)})
  (\bibinfo{year}{2014}), \eprint{1412.7515}.

\bibitem[{\citenamefont{Descotes-Genon
  et~al.}(2013{\natexlab{c}})\citenamefont{Descotes-Genon, Matias, and
  Virto}}]{Descotes-Genon:2013zva}
\bibinfo{author}{\bibfnamefont{S.}~\bibnamefont{Descotes-Genon}},
  \bibinfo{author}{\bibfnamefont{J.}~\bibnamefont{Matias}}, \bibnamefont{and}
  \bibinfo{author}{\bibfnamefont{J.}~\bibnamefont{Virto}},
  \bibinfo{journal}{PoS} \textbf{\bibinfo{volume}{EPS-HEP2013}},
  \bibinfo{pages}{361} (\bibinfo{year}{2013}{\natexlab{c}}),
  \eprint{1311.3876}.

\bibitem[{\citenamefont{Glashow et~al.}(2014)\citenamefont{Glashow, Guadagnoli,
  and Lane}}]{Glashow:2014iga}
\bibinfo{author}{\bibfnamefont{S.~L.} \bibnamefont{Glashow}},
  \bibinfo{author}{\bibfnamefont{D.}~\bibnamefont{Guadagnoli}},
  \bibnamefont{and} \bibinfo{author}{\bibfnamefont{K.}~\bibnamefont{Lane}}
  (\bibinfo{year}{2014}), \eprint{1411.0565}.

\bibitem[{\citenamefont{Altmannshofer
  et~al.}(2014{\natexlab{b}})\citenamefont{Altmannshofer, Gori, Pospelov, and
  Yavin}}]{Altmannshofer:2014pba}
\bibinfo{author}{\bibfnamefont{W.}~\bibnamefont{Altmannshofer}},
  \bibinfo{author}{\bibfnamefont{S.}~\bibnamefont{Gori}},
  \bibinfo{author}{\bibfnamefont{M.}~\bibnamefont{Pospelov}}, \bibnamefont{and}
  \bibinfo{author}{\bibfnamefont{I.}~\bibnamefont{Yavin}},
  \bibinfo{journal}{Phys.Rev.Lett.} \textbf{\bibinfo{volume}{113}},
  \bibinfo{pages}{091801} (\bibinfo{year}{2014}{\natexlab{b}}),
  \eprint{1406.2332}.

\bibitem[{\citenamefont{Mishra et~al.}(1991)}]{Mishra:1991bv}
\bibinfo{author}{\bibfnamefont{S.}~\bibnamefont{Mishra}} \bibnamefont{et~al.}
  (\bibinfo{collaboration}{CCFR Collaboration}),
  \bibinfo{journal}{Phys.Rev.Lett.} \textbf{\bibinfo{volume}{66}},
  \bibinfo{pages}{3117} (\bibinfo{year}{1991}).

\bibitem[{\citenamefont{Ma et~al.}(2002)\citenamefont{Ma, Roy, and
  Roy}}]{Ma:2001md}
\bibinfo{author}{\bibfnamefont{E.}~\bibnamefont{Ma}},
  \bibinfo{author}{\bibfnamefont{D.}~\bibnamefont{Roy}}, \bibnamefont{and}
  \bibinfo{author}{\bibfnamefont{S.}~\bibnamefont{Roy}},
  \bibinfo{journal}{Phys.Lett.} \textbf{\bibinfo{volume}{B525}},
  \bibinfo{pages}{101} (\bibinfo{year}{2002}), \eprint{hep-ph/0110146}.

\bibitem[{\citenamefont{Harigaya et~al.}(2014)\citenamefont{Harigaya, Igari,
  Nojiri, Takeuchi, and Tobe}}]{Harigaya:2013twa}
\bibinfo{author}{\bibfnamefont{K.}~\bibnamefont{Harigaya}},
  \bibinfo{author}{\bibfnamefont{T.}~\bibnamefont{Igari}},
  \bibinfo{author}{\bibfnamefont{M.~M.} \bibnamefont{Nojiri}},
  \bibinfo{author}{\bibfnamefont{M.}~\bibnamefont{Takeuchi}}, \bibnamefont{and}
  \bibinfo{author}{\bibfnamefont{K.}~\bibnamefont{Tobe}},
  \bibinfo{journal}{JHEP} \textbf{\bibinfo{volume}{1403}}, \bibinfo{pages}{105}
  (\bibinfo{year}{2014}), \eprint{1311.0870}.

\bibitem[{\citenamefont{Bell et~al.}(2014)\citenamefont{Bell, Cai, Leane, and
  Medina}}]{Bell:2014tta}
\bibinfo{author}{\bibfnamefont{N.~F.} \bibnamefont{Bell}},
  \bibinfo{author}{\bibfnamefont{Y.}~\bibnamefont{Cai}},
  \bibinfo{author}{\bibfnamefont{R.~K.} \bibnamefont{Leane}}, \bibnamefont{and}
  \bibinfo{author}{\bibfnamefont{A.~D.} \bibnamefont{Medina}},
  \bibinfo{journal}{Phys.Rev.} \textbf{\bibinfo{volume}{D90}},
  \bibinfo{pages}{035027} (\bibinfo{year}{2014}), \eprint{1407.3001}.

\bibitem[{\citenamefont{del Aguila et~al.}(2014)\citenamefont{del Aguila,
  Chala, Santiago, and Yamamoto}}]{delAguila:2014soa}
\bibinfo{author}{\bibfnamefont{F.}~\bibnamefont{del Aguila}},
  \bibinfo{author}{\bibfnamefont{M.}~\bibnamefont{Chala}},
  \bibinfo{author}{\bibfnamefont{J.}~\bibnamefont{Santiago}}, \bibnamefont{and}
  \bibinfo{author}{\bibfnamefont{Y.}~\bibnamefont{Yamamoto}}
  (\bibinfo{year}{2014}), \eprint{1411.7394}.

\bibitem[{\citenamefont{Dumont et~al.}(2014{\natexlab{a}})\citenamefont{Dumont,
  Gunion, Jiang, and Kraml}}]{Dumont:2014wha}
\bibinfo{author}{\bibfnamefont{B.}~\bibnamefont{Dumont}},
  \bibinfo{author}{\bibfnamefont{J.~F.} \bibnamefont{Gunion}},
  \bibinfo{author}{\bibfnamefont{Y.}~\bibnamefont{Jiang}}, \bibnamefont{and}
  \bibinfo{author}{\bibfnamefont{S.}~\bibnamefont{Kraml}},
  \bibinfo{journal}{Phys.Rev.} \textbf{\bibinfo{volume}{D90}},
  \bibinfo{pages}{035021} (\bibinfo{year}{2014}{\natexlab{a}}),
  \eprint{1405.3584}.

\bibitem[{\citenamefont{Dumont et~al.}(2014{\natexlab{b}})\citenamefont{Dumont,
  Gunion, Jiang, and Kraml}}]{Dumont:2014kna}
\bibinfo{author}{\bibfnamefont{B.}~\bibnamefont{Dumont}},
  \bibinfo{author}{\bibfnamefont{J.~F.} \bibnamefont{Gunion}},
  \bibinfo{author}{\bibfnamefont{Y.}~\bibnamefont{Jiang}}, \bibnamefont{and}
  \bibinfo{author}{\bibfnamefont{S.}~\bibnamefont{Kraml}}
  (\bibinfo{year}{2014}{\natexlab{b}}), \eprint{1409.4088}.

\bibitem[{\citenamefont{Aushev et~al.}(2010)\citenamefont{Aushev, Bartel,
  Bondar, Brodzicka, Browder et~al.}}]{Aushev:2010bq}
\bibinfo{author}{\bibfnamefont{T.}~\bibnamefont{Aushev}},
  \bibinfo{author}{\bibfnamefont{W.}~\bibnamefont{Bartel}},
  \bibinfo{author}{\bibfnamefont{A.}~\bibnamefont{Bondar}},
  \bibinfo{author}{\bibfnamefont{J.}~\bibnamefont{Brodzicka}},
  \bibinfo{author}{\bibfnamefont{T.}~\bibnamefont{Browder}},
  \bibnamefont{et~al.} (\bibinfo{year}{2010}), \eprint{1002.5012}.

\bibitem[{\citenamefont{Olive et~al.}(2014)}]{Agashe:2014kda}
\bibinfo{author}{\bibfnamefont{K.}~\bibnamefont{Olive}} \bibnamefont{et~al.}
  (\bibinfo{collaboration}{Particle Data Group}), \bibinfo{journal}{Chin.Phys.}
  \textbf{\bibinfo{volume}{C38}}, \bibinfo{pages}{090001}
  (\bibinfo{year}{2014}).

\bibitem[{\citenamefont{Khachatryan et~al.}(2015)}]{Khachatryan:2015kon}
\bibinfo{author}{\bibfnamefont{V.}~\bibnamefont{Khachatryan}}
  \bibnamefont{et~al.} (\bibinfo{collaboration}{CMS Collaboration})
  (\bibinfo{year}{2015}), \eprint{1502.07400}.

\end{thebibliography}

\end{document}